\documentclass[prd, aps, nofootinbib, preprint, letterpaper]{revtex4-2}

\usepackage{xcolor}
\usepackage[export]{adjustbox}
\usepackage{appendix}
\usepackage{extarrows}
\usepackage{amsmath, amssymb, amsfonts}

\usepackage[hidelinks]{hyperref}
\usepackage{ulem}
\usepackage{setspace}

\newcommand{\checked}[1]{}
\newcommand{\beq}{\begin{equation}}
\newcommand{\eeq}{\end{equation}}
\newcommand{\bqa}{\begin{eqnarray}}
\newcommand{\eqa}{\end{eqnarray}}

\newcommand{\la}{\label}

\newcommand{\wh}{{\hat\omega}}

\begin{document}

\title {Collisional energy loss of a heavy quark in a semiquark-gluon plasma}

\author{Qianqian Du$^{a,b}$, Mudong Du$^{a}$, and Yun Guo$^{a,b,*}$}
\affiliation{
$^a$ Department of Physics, Guangxi Normal University, Guilin, 541004, China\\
$^b$ Guangxi Key Laboratory of Nuclear Physics and Technology, Guilin, 541004, China\\
}

\renewcommand{\thefootnote}{\fnsymbol{footnote}}
\footnotetext[1]{Contact author: yunguo@mailbox.gxnu.edu.cn}
\renewcommand{\thefootnote}{\arabic{footnote}}

\begin{abstract}
By utilizing a background field effective theory, we compute the collisional energy loss of a heavy quark moving through a semiquark-gluon plasma characterized by nontrivial holonomy for Polyakov loops. %By utilizing a background field effective theory which can qualitatively describe the thermodynamics of the quark-gluon plasma above the critical temperature $T_d$, 
We consider the elastic scatterings between the incident heavy quark and the thermal partons with both hard and soft momentum transfers. As compared to the energy loss obtained from the perturbation theory, the hard processes get modified through the thermal distribution functions that depend on the background field, while the proper treatment of the soft processes strongly relies on the use of the hard-thermal-loop resummed gluon propagator derived from the background field effective theory. Our results show that the heavy quark energy loss is significantly suppressed in the semiquark-gluon plasma due to a background field that is self-consistently generated in the effective theory. On the other hand, the suppression has a strong dependence on the temperature of the plasma, which becomes negligible above $2-3 $ times the critical temperature. For a realistic coupling constant, ignoring a relatively weak dependence on the heavy quark velocity, the suppression on the collisional energy loss can be approximated by an overall factor determined solely by the background field. This simple conclusion is expected to be useful for phenomenological applications in the heavy flavor physics.
\end{abstract}

\maketitle
\newpage

\section{Introduction}\la{intro}
Under extremely hot and dense conditions created in the heavy ion experiments, the confined states of hadrons are broken, resulting in a new state of matter known as quark-gluon plasma (QGP)~\cite{Freedman:1976ub,Shuryak:1978ij}. Studying the equation of state is a crucial aspect of understanding the physical properties of the strongly interacting matter. At asymptotically high temperatures, perturbative QCD works very well. Especially, the hard-thermal-loop (HTL) resummed perturbation theory is believed to be a powerful tool, which provides reliable predictions on the thermodynamics of the QGP down to several times the critical temperature $T_d$~\cite{Andersen:2011sf,Haque:2014rua}. On the other hand, the hadron resonance gas model can be used to simulate the confined hadronic phase at temperatures below $T_d$. However, there exists an intermediate region termed as semi-QGP~\cite{Hidaka:2008dr}, from the critical temperature to a few times that, where the physics is of particular interest. For high energy heavy-ion experiments carried out at the Large Hadron Collider (LHC), and especially at Relativistic Heavy Ion Collider (RHIC), the temperatures probed are not far from the critical temperature. Unfortunately, neither the hadron resonance gas nor the HTL resummed perturbation theory can be a reliable theoretical means to explore this intermediate region. 

The partial deconfinement in a semi-QGP is characterized by nontrivial holonomy for Polyakov loops. As the order parameter for the deconfining phase transition in $SU(N)$ gauge theories, the Polyakov loop is nonzero but less than unity in the semi-QGP region according to the lattice simulations~\cite{Gupta:2007ax,Mykkanen:2012ri}. This behavior can be realized by introducing a classical background field for the timelike component of the vector potential $A_0$. The resulting matrix models~\cite{Meisinger:2001cq,Hidaka:2020vna}, which can be considered as a background field effective theory, give rise to a deconfining phase transition through a competition between two different contributions making up the effective potential in this model. One is a perturbative term, favoring the completely deconfined vacuum, and the other is a nonperturbative term, driving the system to confinement. In addition, the predicted thermodynamics is also in a good agreement with the lattice data. Given the success made by the matrix models, it is reasonable to adopt such an effective theory to study some other physical quantities in a semi-QGP. 

In this work, we will focus on the collisional energy loss $-dE/dx$ of an energetic heavy quark moving through a hot and dense QCD plasma which is only partially deconfined. As observed in the relativistic heavy-ion experiments, the suppression of high transverse momentum hadrons can be explained by the energy loss of high energy partons in the plasma. This phenomenon is called the jet quenching, an excellent hard probe which provides a novel window to unravel the fascinating properties of the deconfined nuclear matter~\cite{Wang:1992qdg,Gyulassy:2003mc,Qin:2015srf,Wang:2002ri}. Furthermore, unlike the light quarks which lose energy mainly through the gluon radiation, the collisional energy loss of a heavy quark is comparable to the radiated energy loss because of the dead-cone effect related to the large quark mass~\cite{Dokshitzer:2001zm,Zhang:2003wk,Mustafa:2004dr,Wicks:2005gt}. Therefore, investigating the collisional energy loss induced by the elastic scatterings between the incident heavy quark and thermal medium partons can help us to understand the measured heavy-flavor spectra in nucleus-nucleus collisions. After the first detailed calculation of $-dE/dx$ for a heavy quark in the QCD plasma made by Braaten and Thoma~\cite{Braaten:1991jj,Braaten:1991we}, many attentions have been paid to this research topic, and there were various theoretical developments over the past thirty years; see Refs.~\cite{Romatschke:2003vc,Romatschke:2004au,Herzog:2006gh,Djordjevic:2006tw,Peigne:2007sd,Peigne:2008nd,Lin:2013efa,Elias:2014hua,Han:2017nfz,YousufJamal:2019pen,Peng:2024zvf} and references therein. 

It is certainly interesting to consider the collisional energy loss of a heavy quark by using a background field effective theory when the nonperturbative physics plays an important role and may affect the final results significantly. Our calculation aims to provide quantitative estimates on $-dE/dx$ in a semi-QGP and also to assess the influence of a nonzero background field that is self-consistently generated within this effective theory. A previous work~\cite{Lin:2013efa} also focused on the same problem where the corresponding calculations were carried out in the perturbation theory. However, in the presence of a background field, it is necessary to use an effective theory, which will be further addressed in the following. On the other hand, the main novelty in the current work is that we introduce the HTL resummed gluon propagator for the first time to provide quantitative predictions for the elastic energy loss of a heavy quark in a semi-QGP. This is a crucial step for meaningful computations as elastic scatterings are infrared-sensitive quantities. Such a resummed gluon propagator as derived from the background field effective theory is also important for processes involving soft momentum exchanges, such as the determination of the jet broadening in a semi-QGP~\cite{Moore:2021jwe}.

As a first step towards a full QCD analysis, we only concentrate on a gluonic plasma because including dynamical quarks in the effective theory is currently an open question. The rest of the paper is organized as follows. In Sec.~\ref{defel}, we give the basic definition of the collisional energy loss in QCD and review the theoretical approaches to calculate the hard and soft contributions to $-d E/d x$ in the perturbation theory. In Sec.~\ref{elinsqgp}, after introducing the background field effective theory of a semi-QGP, we discuss how the resummation of the usual hard thermal loops will be modified in the presence of a background field. In addition, the hard and soft contributions to the energy loss are separately derived and the corresponding analytical results are obtained in the weak coupling limit. We show our numerical results of $-d E/d x$ for a charm and bottom quark in Sec.~\ref{nre}, where detailed discussions on the heavy quark velocity/momentum dependence of the collisional energy loss and its suppression in the semi-QGP are also presented. Finally, a summary and outlook can be found in Sec.~\ref{summary}.

\section{The definition of collisional energy loss in QCD}\la{defel}

When a high energy heavy quark propagates through a plasma consisting of light quarks and gluons in thermal equilibrium, it may lose %We set the light quark mass to zero and assume $T\ll m_Q$, $T\ll p$. %and work to leading order in $T/M_Q$ and $T/p$.
energy by interacting with the thermal partons. The rate of energy loss per distance traveled is given by \cite{Braaten:1991jj}
\bqa
-{\frac{dE}{dx}} &=& \frac{1}{v} \int_{m_Q}^{\infty}dE^{'}(E-E^{'})\frac{d\Gamma(E)}{dE^{'}}\, ,
\label{flambda2}
\eqa
where $E$ and ${\bf{v}}={\bf{p}}/E$ denote the energy and velocity of the incident heavy quark with mass $m_Q$. In addition, the interaction rate $\Gamma(E)$ reads
\bqa
\Gamma (E)&=&\frac{1}{2E} \int \frac{d^3 {\bf{p}}'}{(2\pi)^3 2{E}'}\int \frac{d^3 {\bf k}}{(2\pi)^3 2k} n(k)\int \frac{d^3 {\bf {k}}'}{(2\pi)^3 2{k}'}\big(1 - n({k}')\big)\,\nonumber \\
&\times&(2\pi)^4\delta^4(P+K-{P}'-{K}')\,\bigg(\frac{1}{2}\sum_{\mathrm{spins}}\frac{1}{3}\sum_{\mathrm{color}}|\mathcal{M}|^2\bigg)\, .
\label{interaction_rate}
\eqa
In the above equation, $P=(E, {\bf p})$ and $P^\prime=(E^\prime, {\bf p}^\prime)$ are the four-momenta of the incoming and outgoing heavy quark, respectively. The four-momenta of the medium partons, which scatter off of the heavy quark, are denoted by $K=(k, {\bf k})$ and $K^\prime=(k^\prime, {\bf k}^\prime)$. For quark-gluon scattering, the phase space is weighted by a Bose-Einstein distribution $n(k)=(e^{k/T}-1)^{-1}$ and a factor $1-n(k^\prime)$ accounting for the Bose enhancement.

\begin{figure*}[t]
\begin{center}
\includegraphics[width=0.7\linewidth]{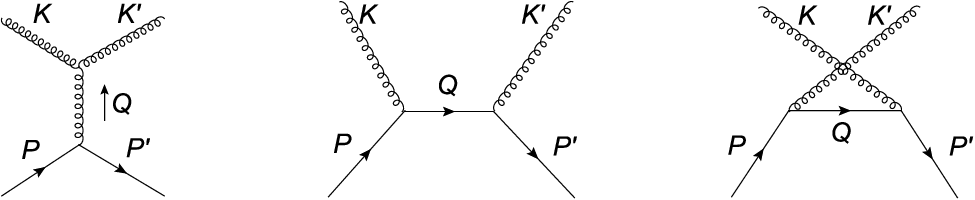}
\end{center}
\vspace{-5mm}
\caption{Tree level Feynman diagrams for the scattering process $Q g\rightarrow Q g$.}
\label{treelevel}
\end{figure*}

The matrix element ${\cal M}$ can be determined by computing the corresponding Feynman diagrams for the elastic scattering processes. It is straightforward to obtain the tree-level contributions as shown in Fig.~\ref{treelevel} where, however, an infrared divergence stemming from the $t$-channel diagram emerges when integrating over the transferred momentum ${\bf q}={\bf p}-{\bf p}^\prime$. Following Braaten and Yuan \cite{Braaten:1991dd}, one can introduce an arbitrary scale $q^*$ for the momentum transferred by which one defines the so-called hard and soft contributions to $- d E/d x$. The hard contributions account for scatterings with momentum transfer larger than $q^*$ and can be expressed as
\bqa
-{\Big(\frac{dE}{dx}\Big)}_{\rm {hard}}^{Qg(t)}&=&\frac{ (4\pi)^3\alpha_s^2}{v} \int \frac{d^3 {\bf k}}{(2\pi)^3} \frac{n(k)}{k}\int \frac{d^3 {\bf {k}}'}{(2\pi)^3 }\frac{1 + n({k}')}{k^\prime} \delta(\omega-{\bf v}\cdot{\bf q})\,\nonumber \\
&\times& \theta(q-q^*)\,\frac{4\omega}{(\omega^2-q^2)^2}\bigg[(k-{\bf v}\cdot {\bf k})^2+\frac{1-v^2}{2}(\omega^2-q^2)\bigg]\, ,
\label{qgtEhard}
\eqa
where $\alpha_s=g^2/(4\pi)$ and $\omega \equiv E -E^\prime$. Unless otherwise stated, we set the number of colors $N=3$ in the following. For the quark-gluon scattering, contributions from the $s$- and $u$-channels are finite for small momentum exchange; therefore, there is no need to introduce the cutoff $q^*$. The corresponding result reads
\beq
-{\Big(\frac{dE}{dx}\Big)}_{\rm {hard}}^{Qg(s+u)}=\frac{(4\pi)^3\alpha_s^2}{2v} \int \frac{d^3 {\bf k}}{(2\pi)^3} \frac{n(k)}{k}\int \frac{d^3 {\bf {k}}'}{(2\pi)^3 }\frac{1 + n({k}')}{k^\prime} \delta(\omega-{\bf v}\cdot{\bf q})\frac{\omega(1-v^2)^2}{(k-{\bf v}\cdot {\bf k})^2}\,.
\label{suE}
\eeq
To obtain the above equations, we consider the thermal partons with typical momentum $k$, $k'\sim T$ and assume the mass and velocity of the heavy quark satisfy $m_Q\gg T$ and $v\gg T/E$. Consequently, contributions suppressed by $T/m_Q$ or $T/p$ have been dropped in $|\mathcal{M}|^2$. Notice that under the above assumptions, interferences between $t$- and $s/u$-channel become negligible and the energy conservation in Eq.~(\ref{interaction_rate}) reduces to $\delta(\omega-{\bf v}\cdot{\bf q})$. Further simplification can be made by realizing the fact that in Eqs.~(\ref{qgtEhard}) and (\ref{suE}), terms involving the product $n_{B}({k})n_{B}({k}')$ are odd under the interchange $\bf{k}\leftrightarrow \bf{k}'$, and thus do not contribute to the energy loss after integrating over the momenta.

The interaction rate $\Gamma$ in Eq.~(\ref{interaction_rate}) can be also expressed in terms of heavy quark self-energy \cite{Braaten:1991jj}. For the soft contribution to $-dE/dx$, the above mentioned  infrared divergence can be eliminated by using the HTL resummed gluon propagator in the calculation of the heavy quark self-energy. It has been shown that \cite{Thoma:2000dx} the resulting soft contribution to $-dE/dx$ is equivalent to the energy loss obtained in the classical plasma physics where $-dE/dx$ is caused by the Lorentz force due to the chromoelectric field induced by the incident heavy quark, namely \cite{Thoma:1990fm,Romatschke:2003vc}
\beq\label{softclassical}
\left(\frac{dE}{dt}\right)_{{\rm soft}}={\rm Re} \int d^3 {\bf x}\,\, {\bf J}^a_{\rm ext}(t, {\bf x}) \cdot {\bf E}^a_{\rm ind}(t, {\bf x})\, ,
\eeq
where the current ${\bf J}^{a}_{\rm ext}(t, {\bf x})={\sf{c}}^a{\bf v}\,\delta^{(3)} ({\bf x}-{\bf v} t)$ represents the propagation of the heavy quark with color charge ${\sf{c}}^a$ defined by $\sum {\sf{c}}^a {\sf{c}}^a=g^2 C_F $. For $SU(3)$, the Casimir invariant $C_F=4/3$.
Using the Vlasov-Maxwell equations, the induced chromoelectric field in the momentum space can be written as
\beq
E^{i,a}_{\rm ind}(Q)=i \omega \big[\Delta^{ij}(Q)-\Delta_0^{ij}(Q)\big] J^{j,a}_{\rm ext}(Q)\,,
\eeq
with $Q=(\omega,{\bf q})$. In the above equation, $\Delta^{ij}(Q)$ denotes the HTL resummed gluon propagator in the temporal axial gauge while $\Delta_0^{ij}(Q)$ is the corresponding temperature-independent free propagator. Explicitly, we have
\bqa
\Delta^{ij}(Q)&=&\frac{1}{\omega^2-q^2-m_D^2 \Pi_T(\wh)}A^{ij}+\frac{1}{\omega^2-m_D^2 \Pi_L(\wh)} B^{ij}\, ,\nonumber \\
\Delta^{ij}_0(Q)&=&\frac{1}{\omega^2-q^2}A^{ij}+\frac{1}{\omega^2} B^{ij}\, ,
\label{normal_gluon_propagator}
\eqa
where the transverse and longitudinal structure functions read
\bqa
\Pi_T(\wh) &=& \frac{\wh^2}{2}\bigg(1-\frac{\wh^2-1}{2\wh}\ln\frac{\wh+1}{\wh-1}\bigg)\,,\nonumber \\
\Pi_L(\wh)& = &  \wh^2 \bigg(-1+\frac{\wh}{2}\ln\frac{\wh+1}{\wh-1}\bigg)\, ,
\label{se}
\eqa
and the two projectors are defined as
\beq
A^{ij}=\delta^{ij}-q^i q^j/q^2\,,\quad\quad\quad B^{ij}=q^i q^j/q^2\,.
\eeq
In addition, $\wh=\omega/q$ and the Debye screening mass $m_D^2= N g^2 T^2/3$. Combining the above equations, we arrive at the following form for the soft contribution to the energy loss:
\bqa
-\bigg(\frac{dE}{dx}\bigg)_{\mathrm{soft}}=\frac{16 \pi\alpha_s}{3 v} {\rm Im}\int\,\frac{d^3 {\bf q}}{(2\pi)^3}({\bf v}\cdot{\bf q}) v^i \big[\Delta^{ij}(Q)-\Delta^{ij}_0(Q)\big]v^j \theta(q^*-q)\Big|_{\omega={\bf v}\cdot{\bf q}} \, ,
\label{soft_def}
\eqa
where the cutoff $q^*$ regulates the logarithmic divergence for large momentum transfer. Such an unphysical behavior appears because Eq.~(\ref{softclassical}) is only valid for soft processes with a typical momentum transfer $q \sim g T$.

\section{The collisional energy loss of a heavy quark in a semi-QGP}
\la{elinsqgp}
In this section, we study the collisional energy loss of a heavy quark in a semi-QGP, where the focus will be put on the influence of a background field ${\cal Q}$ on the corresponding $-dE/dx$. In Sec.~\ref{effth}, we review some basics of the effective theory adopted in our calculation and derive the ${\cal Q}$-dependent resummed gluon propagator. %which is the key ingredient for computing the soft contribution to $-d E/d x$.
The hard and soft contributions to $-dE/dx$ in the presence of a background field will be discussed in Secs.~\ref{hardbf} and \ref{softbf}, respectively.

\subsection{Effective theory of a semi-QGP}\label{effth}

According to the previous discussions, it is possible to introduce a classical background field $A_0^{\rm cl}$ to describe the nontrivial Polyakov loop in the deconfining phase transition for $SU(N)$ gauge theories. The background field is assumed to be constant in spacetime and given by a diagonal matrix in color space,
\beq
(A_0^{\rm cl})_{ab}=\frac{1}{g}{\cal Q}^a \delta_{ab}\, .
\eeq
Furthermore, it satisfies the traceless condition $\sum_{a=1}^{N}{\cal Q}^a=0$ with $a,b=1,\cdots, N$.

The Wilson line in the temporal direction is defined as
\beq
{\bf L}={\cal P}\,{\rm exp}\Big(i g\int^{\beta}_0 A^{\rm cl}_0\, d\tau \Big)\, ,
\eeq
where ${\cal P}$ denotes the time ordering, $\tau$ is the imaginary time and $\beta\equiv1/T$ is the inverse temperature. Accordingly, the gauge invariant Polyakov loop takes the following form:
\beq
\ell=\frac{1}{N}{\rm Tr}\,{\bf L}\, .
\eeq
One can also define higher loops, $(1/N) {\rm Tr}\,{\bf L}^n$, and there are $N-1$ independent loops for $SU(N)$.

After taking into account the classical background field, the effective potential in general $SU(N)$ gauge theories can be computed in the perturbation theory. The one-loop result is
\beq\label{eppt}
{\cal V}_{\rm pt}= \frac{2\pi^2 T^4}{3} \sum_{ab} {\cal P}^{ab,ba} B_4(|{\sf q}^{ab}|)\, .
\eeq
The dependence on the background field ${\cal Q}$ is given by the fourth Bernoulli polynomials $B_4(x)=x^2(1-x)^2-1/30$. For later use, we also need the second Bernoulli polynomials $B_2(x)=x^2-x+1/6$. Notice that the argument $x$ should be understood as $x - [x]$ with $[x]$ the largest integer less than $x$, which is nothing but the modulo function. In the above equation, $a$ and $b$ run from $1$ to $N$ when summing over the color indices. In addition, we define ${\sf q}^{ab} = {\sf q}^{a}-{\sf q}^{b}$ with ${\sf q}^{a} \equiv {\cal Q}^a/(2\pi T)$ and the projection operator $\mathcal{P}^{ab, cd}$ in the double line basis reads
\beq
{\cal P}^{ab,cd}={\cal P}^{ab}_{dc} = \delta^a_d\delta^b_c-\frac{1}{N}\delta^{ab}\delta_{cd}\, .
\eeq
It is more convenient to perform the calculation in the double line basis when a background field is present. In this basis, color indices in the fundamental representation are denoted by $a,b,\cdots = 1,2,\cdots, N$, while in the adjoint representation, they are denoted by a pair of fundamental indices, $ab$. The generators of the fundamental representation are given by $(t^{ab})_{cd}=\frac{1}{\sqrt{2}}\mathcal{P}^{ab}_{cd}$, normalized as ${\rm Tr}\, (t^{ab} t^{cd})=\frac{1}{2} {\cal P}^{ab,cd}$. More details about the double line basis can be found in Refs.~\cite{Hidaka:2009hs,Cvitanovic:1976am}.

The equation of motion for the background field based on Eq.~(\ref{eppt}) indicates that the system is always in a fully deconfined phase with vanishing background field, and no phase transition could happen. It can also be shown that after including the two-loop perturbative corrections~\cite{KorthalsAltes:1993ca,Dumitru:2013xna,Reinosa:2015gxn,Maelger:2017amh,Guo:2018scp}, one still has ${\cal Q} = 0$ as the vacuum. To drive the system to confinement, a nonperturbative contribution needs to be added to the effective potential. It is expected to play an important role near the critical temperature where the confined vacuum is favored. One possible way to generate such a contribution is to include a mass scale in the dispersion relation for gauge bosons, $\omega_k = \sqrt{k^2 + M^2}$, then expand the resulting effective potential in the high temperature limit $M \ll T$. The leading order term in the expansion is nothing but the perturbative ${\cal V}_{\rm pt}$ as given in Eq.~(\ref{eppt}), and the next-to-leading order term appears as the desired nonperturbative contribution in the effective potential which takes the following form~\cite{Meisinger:2001cq}:
\beq\label{epnpt}
{\cal V}_{\rm npt}= \frac{M^2 T^2}{2} \sum_{ab} {\cal P}^{ab,ba}{B_2(|{\sf q}^{ab}|)}\, .
\eeq

An alternative way to generate the above nonperturbative contribution is to embed a two-dimensional ghost field isotropically in four dimensions~\cite{Hidaka:2020vna}. In this case, the mass scale in Eq.~(\ref{epnpt}) should be considered as an upper limit for the transverse momentum of the embedded fields. It indicates that the ghost field is two-dimensional at short distances; however, for distances larger than the scale of confinement, it becomes fully four-dimensional field.

The combination of Eqs.~(\ref{eppt}) and (\ref{epnpt}) leads to an effective theory for studying the semi-QGP. One can derive the equation of motion for the classical background field in this theory, and a nonzero and temperature dependent background field arises. At low temperatures, the solution ${\sf q}^a=(N-2a+1)/(2N)$ minimizes the effective potential,\footnote{This is true under the straight line ansatz where the background fields ${\sf q}^a$ with $a=1,2,\cdots ,N$ have constant spacing and automatically satisfy the traceless condition.} which corresponds to the confining vacuum with vanishing Polyakov loop. When the temperature gets very large, ${\sf q}^a\sim M^2/T^2$ indicates a perturbative vacuum where the Polyakov loop approaches to unity. By requiring the phase transition to occur at the critical temperature $T_d$, the mass scale $M$, which is the only parameter in the effective theory can be determined. For the $SU(3)$ gauge theory, by parametrizing the background field as
\beq\label{su3bf}
{\vec {\sf q}}=({\sf q}, 0, - {\sf q})\, ,
\eeq
the explicit $T$-dependence of the background field in the deconfined phase can be determined from its equation of motion as the following:
\beq\label{bf3}
{\sf q}=\frac{1}{36}\big(9-\sqrt{81-80 (T_d/T)^2}\big)\, ,
\eeq
where $M/T_d =2\sqrt{10} \pi/9$ has been used. %From here on, we use ${\vec {\sf{q}}}$ to denote a set of background fields $q_1,q_2,\cdots, q_N$ for general $SU(N)$. 
It is worth noting that besides the nontrivial behavior of the Polyakov loop in the deconfining phase transition, the above effective theory is also capable of qualitatively reproducing the lattice simulations on the thermodynamics for the $SU(N)$ gauge theories. Further improvements have been proposed, leading to various versions of the matrix models for deconfinement~\cite{Dumitru:2010mj,Dumitru:2012fw,Guo:2014zra,Pisarski:2016ixt}.

In this work, we will investigate the influence of the background field on the heavy quark collisional energy loss when it passes through the hot and dense medium. For the hard scatterings, only thermal distributions of the medium partons %which scatter off of the incident heavy quark 
are affected by the background field. The resulting Bose-Einstein distribution functions are given by~\cite{Furuuchi:2005zp,Hidaka:2008dr}
\begin{equation}\la{bd}
n(k_0,{\sf q}^{ab})=\left\{
\begin{aligned}
&\frac{1}{e^{|k_0|/T- 2 \pi i {\sf q}^{ab}}-1}\equiv n_{+}^{ab}(k_0) \quad\quad {\rm for}\quad k_0>0\\
&\frac{1}{e^{|k_0|/T+2 \pi i {\sf q}^{ab}}-1}\equiv n_{-}^{ab}(k_0) \quad\quad {\rm for}\quad k_0<0 \\
\end{aligned}
\right.\,.
\end{equation}
On the other hand, for scatterings with soft gluon exchange, it becomes necessary to use the HTL resummed gluon propagator which regulates the infrared divergence. In the presence of a background field, the resummed propagator obtained from the Dyson-Schwinger equation gets modified through the ${\cal Q}$-dependent gluon self-energy. 

Within the framework of perturbation theory, it was found that the HTL approximated gluon self-energy is not transverse due to the appearance of an anomalous term at nonzero ${\cal Q}$~\cite{Hidaka:2009hs,Wang:2022dcw}. In addition, this anomalous term also results in an ill-defined gluon resummed propagator~\cite{Guo:2020jvc} as well as an unexpected discontinuity in the free energy at higher order in the coupling constant~\cite{KorthalsAltes:2019yih,KorthalsAltes:2020ryu}. These known issues further justify the necessity of employing an effective theory to properly incorporate the effect of a background field. After taking into account the contributions from the aforementioned two-dimensional ghosts, the anomalous term is completely canceled and the resulting gluon self-energy has the same Lorentz structure as its counterpart at ${\cal Q}=0$, and thus preserves the transversality~\cite{Hidaka:2020vna}.

Based on the Dyson-Schwinger equation, the HTL resummed gluon propagator in the effective theory was first computed in \cite{Guo:2020jvc} where the covariant gauge was used. For our purpose, we carry out a similar calculation in the temporal axial gauge and the $N^2-N$ off diagonal components in color space are given by
\beq
\Delta_{ij}^{ab,cd}(Q,{\vec {\sf{q}}}) \xlongequal[]{a\neq b} \delta^{ad}\delta^{bc}\left[\frac{1}{\omega^2-q^2-
{({\cal M}_D^2)}^{ab}({\vec {\sf{q}}})\Pi_{T}(\wh)}A_{ij}+\frac{1}{\omega^2-{({\cal M}_D^2)}^{ab}({\vec {\sf{q}}})\Pi_{L}(\wh)}B_{ij}\right]\, ,
\label{off_diagonal_propagator}
\eeq
where the ${\cal Q}$-modified screening mass can be written as
\beq\label{offmod}
{({\cal M}_D^2)}^{ab}({\vec {\sf{q}}})= \left[\frac{3}{N}\sum_{e=1}^{N}\bigg(B_2(|{\sf q}^{ae}|)+B_2(|{\sf q}^{eb}|)\bigg)+\frac{3 M^2}{4\pi^2 T^2}\right]m_D^2\, ,
\eeq
with $a\neq b$ and the contribution $\sim M^2$ comes from the two-dimensional ghost field. As compared to Eq.~(\ref{normal_gluon_propagator}), it turns out that the influence of the background field on the transverse and longitudinal gluon self-energy amounts to the same modification on the screening mass. It is simply a ${\cal Q}$-dependent multiplication as given by the terms in the square bracket in Eq.~(\ref{offmod}).

In addition, the diagonal components of the HTL resummed gluon propagator, after multiplying by the projection operator and summing over the color indices, read
\beq
\sum_{ab}{\cal P}^{aa,bb}\Delta_{ij}^{aa,bb}(Q,{\vec {\sf{q}}})=\sum_{\sigma=1}^{N-1}\bigg[\frac{1}{\omega^2-q^2-{({\cal M}_D^2)}^{\sigma}({\vec {\sf{q}}})\Pi_{T}(\wh)}A_{ij}+\frac{1}{\omega^2-{({\cal M}_D^2)}^{\sigma}({\vec {\sf{q}}})\Pi_{L}(\wh)}B_{ij}\bigg]\, .
\label{diagonal_propagator}
\eeq
Similarly, the influence of a nonzero background field is fully encoded in the modified screening masses which can be written as
\beq\label{diamod}
{({\cal M}_D^2)}^{\sigma}({\vec {\sf{q}}}) = \left(1+\frac{3M^2}{4\pi^2 T^2}+\frac{6}{N}\lambda^\sigma\right)m_D^2\, ,
\eeq
where terms in the bracket represent the background field modifications and $\lambda^\sigma$ with $\sigma=1,2,\cdots, N-1$ denote a set of ${\cal Q}$-dependent functions. For $SU(3)$ gauge theory, the explicit form of $\lambda^\sigma$ can be obtained as
\beq
{\vec \lambda}=(3{\sf q}^2-3{\sf q}, \, 9{\sf q}^2-5{\sf q})\, .
\eeq

In fact, only the color sum $\sum_{abcd}{\cal P}^{ab,cd}\Delta_{ij}^{ab,cd}$ is relevant for studying the soft scattering processes. Combining the above results of the diagonal and off diagonal contributions, we arrive at the following compact form:
\beq
\sum_{abcd}{\cal P}^{ab,cd}\Delta_{ij}^{ab,cd}(Q,{\vec {\sf{q}}})=\sum_{ab/\sigma}\bigg[\frac{1}{\omega^2-q^2-{({\cal M}_D^2)}^{ab/\sigma}({\vec {\sf{q}}})\Pi_{T}(\wh)}A_{ij}+\frac{1}{\omega^2-{({\cal M}_D^2)}^{ab/\sigma}({\vec {\sf{q}}})\Pi_{L}(\wh)}B_{ij}\bigg]\, ,
\label{propagator}
\eeq
where $\sum_{ab/\sigma}=\sum_{ab}+\sum_{\sigma}$ denotes a sum over both diagonal and off diagonal gluons and for general $SU(N)$, $a,b=1,2,\cdots, N$ with $a\neq b$ and $\sigma=1,2,\cdots, N-1$.
For vanishing background field, it reduces to\footnote{Temperature independent contributions $\sim M^2$ in Eqs.~(\ref{offmod}) and (\ref{diamod}) are negligible in the high temperature limit where the background field vanishes.}
\beq
\sum_{abcd}{\cal P}^{ab,cd}\Delta_{ij}^{ab,cd}(Q,{\vec {\sf{q}}}) \xlongequal[]{{\cal Q}=0} (N^2-1) \Big[\frac{1}{\omega^2-q^2-m_D^2\Pi_T(\wh)}A^{ij}+\frac{1}{\omega^2-m_D^2\Pi_L(\wh)} B^{ij}\Big]\, ,
\eeq
which indicates that the $N^2-1$ gluons have equal contribution to the above result. For nonzero background field, on the other hand, both the $N^2-N$ off diagonal gluons and $N-1$ diagonal gluons acquire different ${\cal Q}$-dependent modifications, and thus become distinguishable by their associated screening masses. Notice that according to our parametrization Eq.~(\ref{su3bf}), there are only two different screening masses for the six off diagonal gluons in $SU(3)$ gauge theory.

\subsection{The hard scattering processes in the background field }\label{hardbf}

\begin{figure*}[t]
\begin{center}
\includegraphics[width=0.7\linewidth]{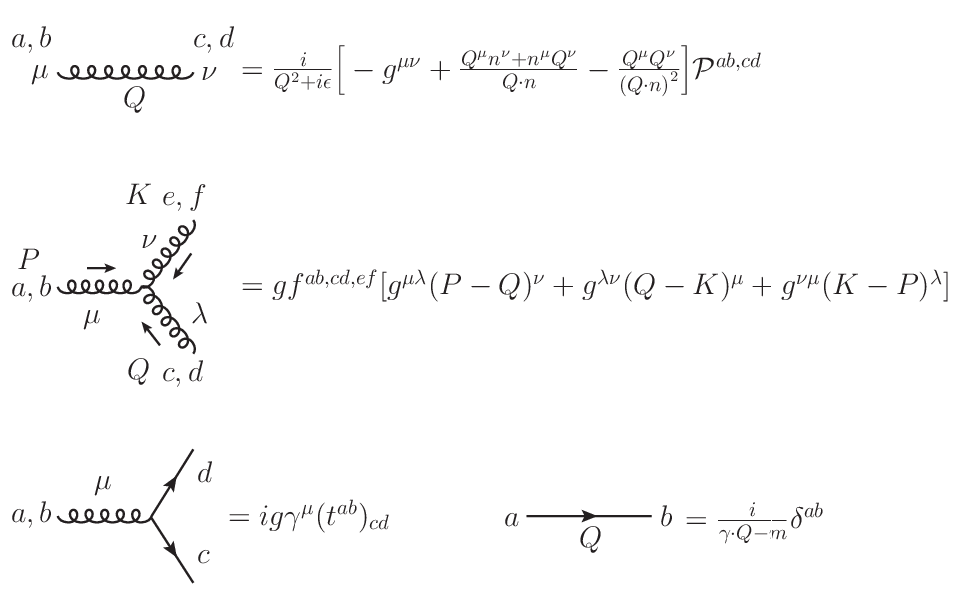}
\end{center}
\vspace{-5mm}
\caption{The Feynman rules in the axial gauge relevant to our calculations. Here, $n^{\mu}$ is a unit vector and the structure constant $f^{ab,cd,ef}=i(\delta^{ad}\delta^{cf}\delta^{eb}-\delta^{af}\delta^{cb}\delta^{ed})/\sqrt{2}$.}
\label{fr}
\end{figure*}

In this subsection, we calculate the hard contributions to the collisional energy loss in the background field. As already mentioned before, the influence of a nonzero ${\cal Q}$ is reflected in the distributions of the thermal partons as given in Eq.~(\ref{bd}). A special feature of such a ${\cal Q}$-dependent distribution is that the gluons are denoted by a pair of the color indices, $ab$. Accordingly, the color structures of the squared matrix element $|\mathcal{M}|^2$ need to be re-computed in the double line basis. We list the corresponding Feynman rules in the axial gauge in Fig.~\ref{fr}.

It is found that for nonvanishing background field, the energy loss $-d E/d x$ due to quark-gluon scattering can be obtained from Eqs.~(\ref{qgtEhard}) and (\ref{suE}) by the replacement,
\bqa
&& 4 \int \frac{d^3 {\bf k}}{(2\pi)^3} \frac{n(k)}{k}\int \frac{d^3 {\bf {k}}^\prime}{(2\pi)^3}\frac{1 - n(k^\prime)}{k^\prime} \rightarrow \nonumber \\
&& \sum_{abcdef} \frac{1}{6} f^{ab,cd,ef} f^{ba,dc,fe} \int \frac{d^3 {\bf k}}{(2\pi)^3} \frac{\big(n_{+}^{ab}(k)+n_{-}^{ab}(k)\big)/2}{k}\int \frac{d^3 {\bf {k}}'}{(2\pi)^3}\frac{1 - \big(n_{
+}^{cd}(k')+n_{-}^{cd}(k')\big)/2}{k^\prime}\, ,\nonumber \\
\eqa
where $4$ is the color factor in zero background field. Analogous to the ${\cal Q}=0$ case, the above form indicates that terms involving the product  $n_{\pm}^{ab}({k})n_{\pm}^{cd}(k^\prime)$ don't contribute to the energy loss. This can be seen by interchanges of the momenta $\bf{k}\leftrightarrow \bf{k}'$ and the color indices $ab\leftrightarrow cd$. Thus, for nonzero background field, we arrive at the following expressions for the hard contributions to $-d E/d x$:
\bqa
-\bigg(\frac{dE}{dx}\bigg)_{\mathrm{BF,hard}}^{Qg(t)} & = &
\frac{2\pi g^4}{v}{\sum_{a,b=1}^3}{\bigg(1-\frac{1}{3}\delta^{ab}\bigg)}\int\frac{d^3\mathbf{k}}{(2\pi)^3 k}\frac{n_{+}^{ab}(k)+n_{-}^{ab}(k)}{2}\int\frac{d^3\mathbf{{k}'}}{(2\pi)^3 {k}'}\delta(\omega-\mathbf{v\cdot q})\nonumber\\ &\times&
 \frac{\omega}{(\omega^2-q^2)^2}\left[ (k-\mathbf{v\cdot k})^2+\frac{1-v^2}{2}(\omega^2-q^2) \right]\theta(q-q^*)\, ,
\label{qgtEhardBF}
\eqa
\bqa
-\bigg(\frac{dE}{dx}\bigg)_{\mathrm{BF,hard}}^{Qg(s+u)}&=&\frac{\pi g^4}{4v}\sum_{a,b=1}^3\bigg(1-\frac{1}{3}\delta^{ab}\bigg)\int\frac{d^3\mathbf{k}}{(2\pi)^3 k}\frac{n_{+}^{ab}(k)+n_{-}^{ab}(k)}{2}\nonumber \\
&\times&\int\frac{d^3\mathbf{{k}'}}{(2\pi)^3 {k}'} \delta(\omega-\mathbf{v\cdot q})\frac{\omega (1-v^2)^2}{(k-\mathbf{v\cdot k})^2}\,.
\label{qgsuEhardBF}
\eqa
The energy loss due to the $s$- and $u$-channel scatterings can be obtained analytically, which reads
\beq
-\bigg(\frac{dE}{dx}\bigg)_{\mathrm{BF,hard}}^{Qg(s+u)}=\alpha_s^2 \pi T^2 {\sum_{ab}}{\bigg(1-\frac{1}{3}\delta^{ab}\bigg)} B_2(|{\sf q}^{ab}|)f_1(v)=\frac{\alpha_s^2 T^2}{\pi} \sum_{n=1}^{\infty} \frac{|{\rm Tr}\,{\bf L}^n|^2-1}{n^2} f_1(v)\, ,
\label{hardBFsimply}
\eeq
where
\bqa
f_1(v)&=&\frac{1}{v}-\frac{1-v^2}{2v^2} \ln \frac{1+v}{1-v}\, .
\eqa
To get the above result, we used
\beq\label{bain}
\sum_{ab} \int_0^\infty k\, {n_{\pm}^{ab}(k)} d k= T^2 \sum_{n=1}^{\infty}\frac{|{\rm Tr}\,{\bf L}^n|^2}{n^2} =\pi^2 T^2 \sum_{ab}  B_2(|{\sf q}^{ab}|)\,.
\eeq
Our results show that for the $s$- and $u$-channel scatterings, the modification of a background field is simply given by the factor $\sum_{ab} (1-\delta^{ab}/3) B_2(|{\sf q}^{ab}|)$. In the case of $SU(3)$, it equals $4(1-3{\sf q})^2/3$, which reduces to $4/3$ for vanishing background field as expected. However, integrals appearing in the $t$-channel scattering turn out to be more involved. To proceed further, we make the following change of the integral variables:
\bqa\label{changeva}
\frac{1}{2(2\pi)^2}\int \frac{d^3 \mathbf{k}}{k} \int \frac{d^3\mathbf{k}'}{k^\prime}&\to& \int_{\frac{1+v}{2}{q^*}}^{\infty} d k\int_{ q^*}^{\frac{2k}{1+v}} q dq \int_{-v q}^{v q} d\omega
+\int_{\frac{1+v}{2}{ q^*}}^{\infty} d k\int_{\frac{2k}{1+v}}^{\frac{2k}{1-v}}q dq \int_{q-2k}^{v q} d\omega
\nonumber \\&+&\int_{\frac{1-v}{2}{q^*}}^{\frac{1+v}{2}{q^*}} d k\int_{q^*}^{\frac{2k}{1-v}} q dq \int_{q-2k}^{v q} d\omega\,.
\eqa
The above integral intervals are determined to meet the constraint condition that $-1\le \cos \theta \le 1$, where $\theta$ denotes the angle between $\mathbf{k}$ and $\mathbf{k}'$ and $\cos \theta= (k^2+(\omega+k)^2-q^2)/(2k(k+w))$. Further constraints due to the two theta functions $\theta(q-q^*)$ and $\theta(v^2 q^2-\omega^2)$ have also been taken into account in Eq.~(\ref{changeva}). The second theta function arises after averaging the integrand over the direction of ${\bf v}$, which is valid because the energy loss is independent of the incident direction of the heavy quark in an isotropic medium.

The cutoff $q^*$ was introduced to eliminate the infrared divergence associated with soft momentum transfer. In fact, in the limit $q^* \rightarrow 0$, only the first integral over $q$ in Eq.~(\ref{changeva}) becomes divergent, and thus needs to be regulated. Consequently, a simplified version of Eq.~(\ref{changeva}) can be obtained as
 \beq\label{changevav2}
\frac{1}{2(2\pi)^2}\int \frac{d^3\mathbf{k}}{k} \int \frac{d^3\mathbf{k}'}{k^\prime}\to \int_{0}^{\infty} d k\int_{ q^*}^{\frac{2k}{1+v}} q dq \int_{-v q}^{v q} d\omega
+\int_{0}^{\infty} d k\int_{\frac{2k}{1+v}}^{\frac{2k}{1-v}} q dq \int_{q-2k}^{v q} d\omega \,,
\eeq
which coincides with the variable change as used in \cite{Braaten:1991jj} and leads to an analytical form for the energy loss even for nonzero background field. The result can be expressed as
\bqa\label{hqgt}
-\bigg(\frac{dE}{dx}\bigg)_{\mathrm{BF,hard}}^{Qg(t)}&=&\frac{2\alpha_s^2}{\pi}{\sum_{ab}} {\bigg(1-\frac{1}{3}\delta^{ab}\bigg)} \int_0^\infty k dk\, \frac{n_{+}^{ab}(k)+n_{-}^{ab}(k)}{2} \bigg[ f_2(v)+ f_1(v)\nonumber \\
&\times&  \Big(\ln(k/q^{*}) -\frac{1}{2} \Big) \bigg]\nonumber \\&=& \frac{2\alpha_s^2 T^2}{\pi} \sum_{n=1}^{\infty} \frac{|{\rm Tr}\,{\bf L}^n|^2-1}{n^2} \bigg[ f_1(v)  \bigg(\ln \frac{T}{q^{*}}+\frac{1}{2} -\gamma-\ln n \bigg)+ f_2(v)\bigg]\,,
\eqa
where
\bqa
f_2(v)&=& \Big[\frac{1}{2}-\frac{1}{2}\ln (1-v^2)+ \ln 2 \Big]f_1(v)-\frac{1-v^2}{4v^2}\Big[{\rm Li}_2\big(\frac{1+v}{2}\big)\nonumber \\&-&{\rm Li}_2\big(\frac{1-v}{2}\big)+\frac{1}{2}\ln\frac{1-v^2}{4} \ln\frac{1+v}{1-v}\Big]-\frac{2}{3} v\, ,
\eqa
and ${\rm Li}_2(x)$ is the polylogarithm. As we can see in the limit $q^* \rightarrow 0$, the above result has a logarithmic divergence $\sim \ln q^*$ and the background field modification on this divergent part is the same as that on the $s$- and $u$-channel contributions. On the other hand, for the finite part, besides the term $\sim k$ in the integral over $k$, there is also a term $\sim k \ln k$, which leads to a more involved ${\cal Q}$ dependence of the energy loss which cannot be described simply by the Bernoulli polynomial.

It can also be shown that the hard contribution in Eq.~(\ref{qgtEhardBF}) is always positive when varying $q^*$. %if eq.~(\ref{changeva}) is used to perform the integrals.
However, Eq.~(\ref{hqgt}) may lead to a negative energy loss at large $q^*$ because of the simplification used in Eq.~(\ref{changevav2}). Therefore, Eq.~(\ref{changevav2}) is only a good approximation when $q^* \ll T$. For an arbitrary cutoff, one should adopt Eq.~(\ref{changeva}) to change the integral variables for mathematical rigour. Thus, Eq.~(\ref{qgtEhardBF}) can be written as
\bqa
-\bigg(\frac{dE}{dx}\bigg)_{\mathrm{BF,hard}}^{Qg(t)} & = &
\frac{3\alpha_s^2}{\pi v^2}{\sum_{ab}}{\bigg(1-\frac{1}{3}\delta^{ab}\bigg)} \iiint_{{\rm sum}} d k\,  dq\,   d\omega \frac{n_{+}^{ab}(k)+n_{-}^{ab}(k)}{2} \frac{\omega}{q^2}    \nonumber\\ &\times& \bigg[  \frac{k^2+ k\omega+\omega^2/4}{q^2}-\frac{1-v^2}{q^2-\omega^2}\frac{k^2+k \omega+q^2}{3}-\frac{v^2}{12}\bigg]\, ,
\label{qgtEhardBFfinal}
\eqa
where we use the shorthand notation $\iiint_{{\rm sum}}  d k\, d q\, d \omega$ to denote the sum of the three triple integrals as given in Eq.~(\ref{changeva}). Consequently, the integral over $k$ needs to be carried out numerically. Similar to the ${\cal Q}=0$ case discussed in \cite{Romatschke:2003vc}, the energy loss based on Eq.~(\ref{qgtEhardBFfinal}) is always positive for an arbitrary cutoff and agrees with Eq.~(\ref{hqgt}) in the small $q^*$ region.

\subsection{The soft scattering processes in the background field }\label{softbf}

The soft contributions to the collisional energy loss can be studied based on Eq.~(\ref{soft_def}), where the influence of the background field amounts to a modification on the gluon propagator. In the double line basis, the gluon propagator is not diagonal in the color space and summing over the color indices becomes nontrivial. In order to explicitly show the color structures, we rewrite Eq.~(\ref{soft_def}) as
\bqa\label{defsoftbf}
-\bigg(\frac{dE}{dx}\bigg)_{\mathrm{BF, soft}} &=&\sum_{abcd} \frac{4\pi \alpha_s}{v} \frac{1}{2 N} {\cal P}^{ab,cd}\, {\rm Im}\int\frac{d^3\mathbf{q}}{(2\pi)^3}(\mathbf{v}\cdot\mathbf{q})v^i\big[\Delta_{ij}^{ab,cd}(Q, {\sf q})\nonumber \\
&-&{(\Delta_0)}_{ij}^{ab,cd}(Q)\big]v^j \theta(q^*-q)\Big|_{\omega={\bf v}\cdot{\bf q}}\, .
\eqa
In the above equation, ${(\Delta_0)}_{ij}^{ab,cd}(Q)$ is not the usual bare propagator in perturbation theory. It actually corresponds to the zero temperature limit of the resummed propagator in the effective theory and can be written as
\beq
{(\Delta_0)}_{ij}^{ab,cd}(Q)= {\cal {P}}^{ab,cd}\bigg[\frac{1}{\omega^2-q^2-
 \frac{3 g^2 M^2 }{4\pi^2}\Pi_{T}(\wh)}A_{ij}+\frac{1}{\omega^2- \frac{3 g^2  M^2}{4\pi^2}\Pi_{L}(\wh)}B_{ij}\bigg]\, .
\label{newbarepro}
\eeq
The usual bare propagator can be recovered by setting $M^2=0$ in the above equation which has no contribution to the energy loss because of a vanishing imaginary part. However, the propagator as given in Eq.~(\ref{newbarepro}) leads to an extra contribution which as we will see, is important to ensure the cancellation of the cutoff dependence in the weak coupling limit although such a contribution is temperature independent.

In the limit ${\cal Q}\rightarrow 0$, the resummed gluon propagation $\Delta_{ij}^{ab,cd}(Q, {\sf q})$ is also proportional to ${\cal P}^{ab,cd}$, and the sum over the color indices in Eq.~(\ref{defsoftbf}) can be easily carried out, which leads to the following color factor:
\beq
\sum_{abcd}\frac{1}{2 N} {\cal P}^{ab,cd} {\cal P}^{ab,cd}=C_F=\frac{4}{3}\, ,
\eeq
for $SU(3)$. On the other hand, when ${\cal Q}\neq 0$, based on the resummed gluon propagator as given in Sec.~\ref{effth}, the following compact form can be found for the collisional energy loss due to soft scatterings,
\bqa
-\bigg(\frac{dE}{dx}\bigg)_{\mathrm{BF,soft}}
&=& \frac{\alpha_s}{12\pi v^2}\sum_{ab/\sigma/s}\mathrm{Im} \int^{+v}_{-v}d\hat{\omega}\frac{v^2-\hat{\omega}^2}{\hat{\omega}}\bigg[\frac{\hat{\omega}^2}{(1-\hat{\omega}^2)^2} F^{ab/\sigma/s}_{T}(\hat{\omega},{\vec {\sf{q}}})\nonumber \\
&\times&\ln\frac{(1-\hat{\omega}^2)(q^*)^2+ F^{ab/\sigma/s}_{T}(\hat{\omega},{\vec {\sf{q}}})}{F^{ab/\sigma/s}_{T}(\hat{\omega},{\vec {\sf{q}}})}+\frac{1}{v^2-\hat{\omega}^2}F^{ab/\sigma/s}_{L}(\hat{\omega},{\vec {\sf{q}}})\nonumber \\
&\times&  \ln\frac{F^{ab/\sigma/s}_{L}(\hat{\omega},{\vec {\sf{q}}})-(q^*)^2\hat{\omega}^2}{F^{ab/\sigma/s}_{L}(\hat{\omega},{\vec {\sf{q}}})}\bigg]\, ,
\label{offdiagsoft}
\eqa
where we used $d^3\mathbf{q}=2\pi q^2 dq d {\hat \omega}/v$ and the integral over $q$ has been done analytically. Furthermore, the shorthand notation of the color sums is defined as
\beq\label{sum}
\sum_{ab/\sigma/s}=\sum_{ab}+\sum_{\sigma}-\sum_{s}\, ,
\eeq
where the first two sums on the right-hand side have already appeared in Eq.~(\ref{propagator}). Notice the minus sign for the last sum which is related to the contribution associated with ${(\Delta_0)}_{ij}^{ab,cd}(Q)$ in Eq.~(\ref{defsoftbf}). In the high temperature limit where ${\cal Q}\rightarrow 0$, such a contribution becomes negligible as compared to the contributions from the first two sums. However, near the critical temperature, the three sums in Eq.~(\ref{sum}) give comparable contributions to the energy loss. In addition, we have
\beq
F^{ab/\sigma/s}_{T/L}(\hat{\omega},{\vec {\sf{q}}})={({\cal M}_D^2)}^{ab/\sigma/s}({\vec {\sf{q}}})\Pi_{T/L}(\hat{\omega})\, ,
\eeq
where $({\cal M}_D^2)^{s}=3 g^2 M^2/(4\pi^2)$ is independent on the color index $s$ and for $SU(N)$, $s=1,2,\cdots, N^2-1$. Therefore, the purpose of introducing such a trivial sum over $s$ is merely to make the form of Eq.~(\ref{offdiagsoft}) compact.

By assuming $q^* \gg g T \sim m_D$, one can expand the logarithm in Eq.~(\ref{offdiagsoft}). Keeping only the leading order terms in the expansion, an analytical expression for the $q^*$-dependent part can be obtained which in the case of $SU(3)$ reads
\beq
\frac{\alpha_s}{12}\sum_{ab/\sigma/s} {({\cal M}_D^2)}^{ab/\sigma/s} f_1(v) \ln q^* = \frac{8 \pi \alpha_s^2 T^2}{3} f_1(v) (1-3 {\sf q})^2 \ln q^*\, .
\eeq
Furthermore, the $q^*$-dependent part of the hard contribution in Eq.~(\ref{hqgt}) can be written as
\bqa
-\frac{2\alpha_s^2 T^2}{\pi} \sum_{n=1}^{\infty}\frac{|{\rm Tr}\,{\bf L}^n|^2-1}{n^2} f_1(v) \ln q^*
&=& -\frac{8 \pi \alpha_s^2 T^2}{3} f_1(v) (1-3 {\sf q})^2 \ln q^*\, .
\eqa
Clearly, the cutoff dependence cancels between the hard and soft contributions, which is similar to the case of zero background field as studied in Ref.~\cite{Braaten:1991jj}. Notice that the above cancellation does not depend on the details of the background field, namely, a specified temperature dependence as determined by Eq.~(\ref{bf3}) is not needed to make the cancellation happen. Here, contributions from the last sum in Eq.~(\ref{sum}) are important to ensure the exact cancellation. 

One can also get the full leading order result of the soft contributions when expanding Eq.~(\ref{offdiagsoft}) with $q^*\gg g T$, which is given by
\bqa\label{exsoft}
-\bigg(\frac{dE}{dx}\bigg)_{\mathrm{BF,soft}}
&=& \frac{8 \pi \alpha_s^2 T^2}{3} (1 -3 {\sf q})^2\Big[f_1(v) \ln \big(q^*/m_D\big) - f_3(v)\Big]\nonumber \\
&-& \frac{\alpha_s}{24}f_1(v)\sum_{ab/\sigma/s} {({\cal M}_D^2)}^{ab/\sigma/s}\ln \big({({\cal M}_D^2)}^{ab/\sigma/s}/m_D^2\big)\,.
\eqa
In the high temperature limit where ${\sf q} \rightarrow 0$, the second line in the above equation can be neglected, and we get the corresponding result at zero background field~\cite{Braaten:1991we}
\beq
-\bigg(\frac{dE}{dx}\bigg)_{\mathrm{BF=0,soft}}
= \frac{8 \pi \alpha_s^2 T^2}{3} \Big[f_1(v) \ln \big(q^*/m_D\big) - f_3(v)\Big]\, ,
\eeq
where
\beq
f_3(v)=\frac{1}{v^2}\int_{0}^{v} d \wh\,  \wh^2 \Big[\frac{v^2-\wh^2}{2(1-\wh^2)^2} \big(G_T(\wh)+\ln (\pi \wh /4)\big)+\big(G_L(\wh)+\ln (\pi \wh /2)\big)\Big]\, ,
\eeq
with
\beq
G_{T/L}(\wh)=Q_{T/L}(\wh) \Big[\frac{\pi}{2}-\arctan \big(Q_{T/L}(\wh)\big)\Big]+\frac{1}{2}\ln \big(1+Q_{T/L}^2(\wh)\big) \, ,
\eeq
and
\beq
Q_{T}(\wh)=\frac{1}{\pi} \Big[\ln\big(\frac{1+\wh}{1-\wh}\big)+\frac{2\wh}{1-\wh^2}\Big]\,,\quad\quad Q_{L}(\wh)=\frac{1}{\pi} \Big[\frac{2}{\wh}-\ln\big(\frac{1+\wh}{1-\wh}\big)\Big]\,.
\eeq
The above result also indicates that the background field modification on the soft contributions is not simply an overall factor $(1-3{\sf q})^2$ due to the appearance of the second line in Eq.~(\ref{exsoft}). Similar behavior has already been found in the $t$-channel contribution in the hard scatterings; see Eq.~(\ref{hqgt}). 

We should point out that the simplification used in Eq.~(\ref{changevav2}) is valid when $q^*\ll T$, while the expansion of Eq.~(\ref{offdiagsoft}) is ensured to be convergent if $q^*\gg g T$. Therefore, the above cancellation of the cutoff dependence requires a very small coupling constant in order to meet the assumption $g T\ll q^* \ll T$ which is also the prerequisite to use the energy loss obtained in Eqs.~(\ref{hqgt}) and (\ref{exsoft}). Similar to the negative energy loss obtained from Eq.~(\ref{hqgt}) for very large $q^*$, Eq.~(\ref{exsoft}) also leads to a negative result when the cutoff becomes too small as compared to the screening mass. On the other hand, when extrapolating to a realistic value of the coupling constant, the above assumption cannot be well satisfied. Consequently, for an arbitrary cutoff, we need to use Eq.~(\ref{changeva}) to carry out the integrals for the hard contributions and evaluate the soft contributions by Eq.~(\ref{offdiagsoft}) without any expansion. The total energy loss, which needs to be determined  numerically, does show a $q^*$ dependence. Such an ambiguity can be eliminated by using the principle of minimum sensitivity~\cite{Romatschke:2003vc}, and thus the cutoff $q^*$ is fixed by minimizing the total energy loss with respect to $q^*$. 

Finally, let us focus on the ${\cal Q}$ dependence of the collisional energy loss based on our results in the weak coupling limit. For the hard processes, since $q^* \ll T $, the dominant $t$-channel contribution comes from the term $\sim \ln (T/q^*)$ in Eq.~(\ref{hqgt}). Therefore, the influence of the background field amounts to a ${\sf q}$-dependent factor reading $(1 -3 {\sf q})^2$ for $SU(3)$.\footnote{The same is true for the $s$- and $u$-channel contributions. However, there is no logarithmic enhancement in Eq.~(\ref{hardBFsimply}), therefore, as compared to the $t$-channel contribution, contributions from $s$- and $u$-channels can be neglected.} For the soft processes, keeping only the logarithmically enhanced contribution $\sim \ln(q^*/m_D)$ with $q^* \gg m_D$, the same conclusion can be drawn from Eq.~(\ref{exsoft}). As  a result, the $q^*$-independent total energy loss equals the factor $(1 -3 {\sf q})^2$ times the corresponding result at vanishing background field. In the weak coupling limit, the dominant contribution in the total energy loss is $\sim \ln (1/g)$ which comes from the soft processes. Therefore, a proper treatment on the soft momentum exchange with the hard thermal loop resummation is necessary.

\section{Numerical results and discussions}\label{nre}

In this section, we present our numerical results for the collisional energy loss of a heavy quark in a semi-QGP which consists of the hard contributions from Eqs.~(\ref{hardBFsimply}) and (\ref{qgtEhardBFfinal}) as well as the soft contributions from Eq.~(\ref{offdiagsoft}). As already mentioned above, the cutoff dependence of the total energy loss can be fixed through the following equation:
\beq
-\frac{d}{dq^{*}}\bigg({\frac{dE}{dx}}\bigg)_{\mathrm {BF}}{\bigg|}_{q^{*}=q^{\rm{pms}}}=0\, ,\quad {\rm with}\quad -\bigg(\frac{dE}{dx}\bigg)_{\mathrm {BF}}\equiv -\bigg(\frac{dE}{dx}\bigg)_{\mathrm{BF, soft}}-{\Big(\frac{dE}{dx}\Big)}_{\rm{BF, hard}}\, ,
\label{variation}
\eeq
and thus, the resulting $-d E/ d x$ serves as a lower bound for the heavy quark energy loss. We consider the collisional energy loss above the critical temperature $T_d=270\,{\rm MeV}$ for the $SU(3)$ gauge theory. Like the zero background field case, our numerical evaluations suggest that the energy loss shows a weak $q^*$ dependence as we decrease the coupling constant and eventually the energy loss coincides with the result given by the sum of Eqs.~(\ref{hardBFsimply}), (\ref{hqgt}) and (\ref{exsoft}) in the weak coupling limit where the $q^*$ dependence vanishes. For simplicity, we fix the coupling constant $\alpha_s=0.3$, corresponding to its typical value in the semi-QGP region.

\begin{figure*}[t]
\begin{center}
\includegraphics[width=0.45\linewidth]{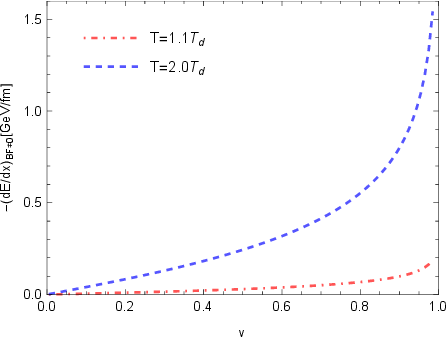}
\includegraphics[width=0.45\linewidth]{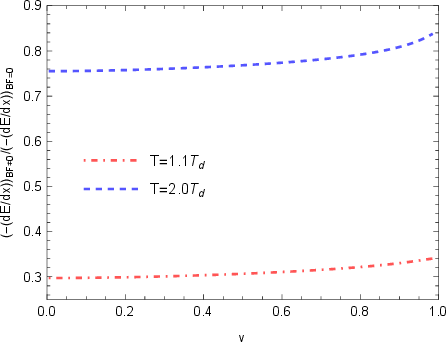}
\end{center}
\vspace{-5mm}
\caption{Collisional energy loss in a background field (left) and the ratio of the energy loss with and without the background field (right) as a function of the heavy quark velocity $v$ at $T=1.1T_d$ and $T=2.0T_d$.}
\label{elvsv}
\end{figure*}

Given the temperature dependence of the background field in Eq.~(\ref{bf3}), the collisional energy loss depends only on the temperature of the plasma $T$ and the velocity of the incident heavy quark $v$. In the left plot of Fig.~\ref{elvsv}, we show $-d E/d x$ as a function of $v$ at different temperatures. As we can see, the collisional energy loss increases with increasing velocity and a very rapid growth is observed in the large $v$ region. Qualitatively, this coincides with the behavior as observed at vanishing background field.\footnote{The results without considering the background field correspond to those obtained in the perturbation theory. Besides setting ${\cal Q}=0$, the mass scale $M^2$ in Eqs.~(\ref{offmod}) and (\ref{diamod}) is also zero, namely, the $N^2-1$ gluons have the same screening mass $m_D$ for $SU(N)$.} It is also found that the heavy quark energy loss is very sensitive to the temperature of the plasma, a significant enhancement appears when the medium moves from a region close to the critical temperature $T_d$ into the perturbative deconfining region, where we choose $T/T_d= 2.0$, corresponding to a rather small background field according to Eq.~(\ref{bf3}). A more direct way to see the influence of the background field is to consider the ratio of the energy loss with and without ${\cal Q}$ which can be found in the right plot of Fig.~\ref{elvsv}. In general, the collisional energy loss is reduced in the presence of a background field. For a given temperature, this ratio, or in other words, the suppression of the energy loss is insensitive to the velocity as long as $v$ is not very large. Therefore, considering small and intermediate incident velocities, the collisional energy loss in a background field can be well approximated with the corresponding energy loss at ${\cal Q}=0$ multiplied by an overall $T$-dependent factor. %Roughly speaking, the background field modification on $(-d E/ d x)_{{\rm BF=0}}$ is thus simply given by such a factor. 
Notice that without considering the background field, the temperature dependence of the energy loss is very simple, {\em i.e.}, $- d E/ d x \sim T^2$, when a fixed coupling is used. This is obviously not true for nonzero background field because the above mentioned factor changes with $T$ significantly. On the other hand, a notable increase of the energy loss ratio can be found in the large $v$ region.\footnote{In fact, our numerical results show that the energy loss ratio has a turning point at a velocity very close to $1$, which is not visible in Fig.~\ref{elvsv}. However, in the ultrarelativistic limit, finite quark mass corrections become important there and for a realistic charm or bottom quark, calculation beyond our approximation is needed which we will discuss in the following.}

\begin{figure*}[t]
\begin{center}
\includegraphics[width=0.48\linewidth]{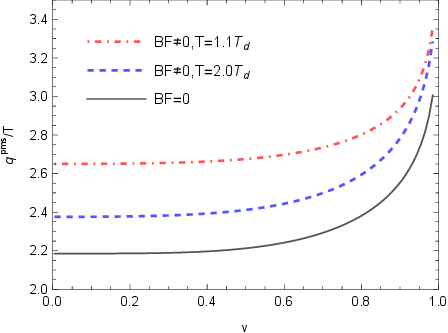}
\end{center}
\vspace{-5mm}
\caption{The velocity dependence of $q^{\rm {pms}}/T$ with and without the background field.}
\label{qsvsv}
\end{figure*}

In Fig.~\ref{qsvsv}, we show the velocity dependence of the cutoff $q^{\rm pms}$ as determined by Eq.~(\ref{variation}). No matter if a background field is taken into account or not, at a given temperature, $q^{\rm pms}$ exhibits a very weak $v$-dependence for small and intermediate velocities. However, a large increase of $q^{\rm pms}$ is found when the velocity gets large. Furthermore, for nonzero background field, $q^{\rm {pms}}/T$ changes when varying the temperature $T$ at a fixed velocity. This is very different from the ${\cal Q}=0$ case where $q^{\rm {pms}}/T$ becomes a constant. We can expect that in the high temperature limit, $q^{\rm {pms}}/T$ would be identical to its counterpart at vanishing background field. Therefore, the background field has a significant impact on the cutoff near the deconfinement point, which gradually diminishes as the temperature increases.

It should be noted that in principle the above results are applicable for infinitely large quark mass because of the approximations used in our calculation. For a charm and bottom quark, finite mass corrections become important for very small ($v \to 0$) and large ($v \to 1$) velocities. On the one hand, when the incident velocity is smaller than the thermal velocity $\sim (T/m_Q)^{1/2}$, the heavy quark is expected to gain energy from a collision. As a result, $-dE/dx$ must change sign at sufficiently small $v$~\cite{Thoma:1990fm}. %some small $v$ which was estimated to be $\sim \sqrt{3 T/m_Q}$ 
On the other hand, a more accurate treatment on the energy loss is required for ultrarelativistic energies $E \gg m_Q^2/T$ where the largest momentum transfer used in Eq.~(\ref{changeva}) is no longer valid,\footnote{The largest momentum transfer $2k/(1-v)$ in Eq.~(\ref{changeva}) is only an approximation, which becomes exact when $E \ll m_Q^2/T$.} and some dropped terms in the squared matrix element may also contribute significantly~\cite{Peigne:2007sd}. A crossover energy was estimated to be $\sim 1.8 m_Q^2/T$ in Ref.~\cite{Braaten:1991dd} for QCD at vanishing background field. Accordingly, there exists an upper bound of the heavy quark velocity given by $ 1-0.28(T/m_Q)^2$, above which finite mass corrections start to  play a role. However, with the typical temperatures achieved in the high energy heavy-ion experiments, we can expect that for both charm and bottom quark our results are reliable up to very large incident velocities.

\begin{figure*}[t]
\begin{center}
\includegraphics[width=0.45\linewidth]{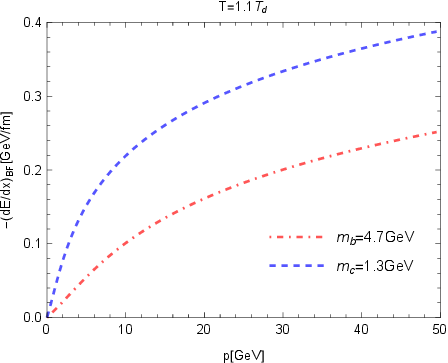}
\includegraphics[width=0.45\linewidth]{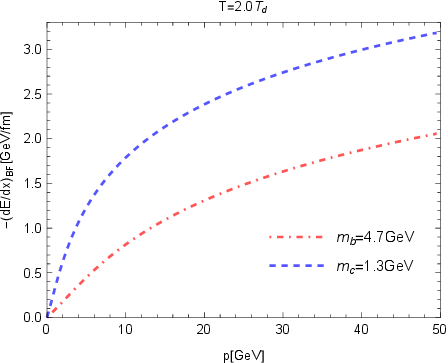}
\end{center}
\vspace{-5mm}
\caption{Energy loss of the bottom ($m_b=4.7\,{\rm GeV}$) and charm quark ($m_c=1.3\,{\rm GeV}$) in a background field as a function of momentum $p$ at $T=1.1T_d$ (left), and $T=2.0T_d$ (right).}
\label{elvsp}
\end{figure*}

\begin{figure*}[t]
\begin{center}
\includegraphics[width=0.45\linewidth]{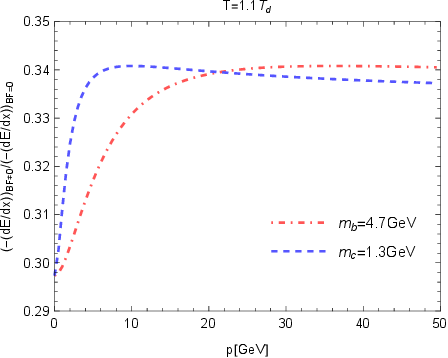}
\includegraphics[width=0.45\linewidth]{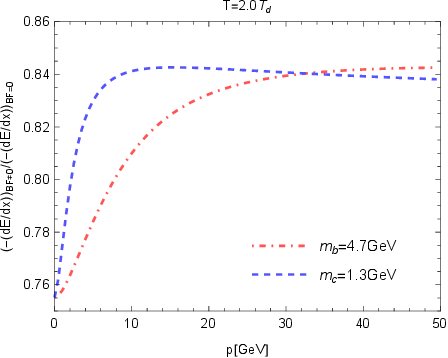}
\end{center}
\vspace{-5mm}
\caption{The ratio of the energy loss with and without the background field as a function of the heavy quark momentum $p$ for charm and bottom at $T=1.1T_d$ (left) and $T=2.0T_d$ (right).}
\label{ravsp}
\end{figure*}

In order to see the flavor dependence of the collisional energy loss, we show in Fig.~\ref{elvsp} $-d E/d x$ for the charm and bottom quark as a function of the momentum $p$ at different temperatures. The momentum $p$ of the heavy quark can be related to its velocity through the equation $p=v m_Q/\sqrt{1-v^2}$, and we choose $m_c=1.3\,{\rm GeV}$ and $m_b=4.7\,{\rm GeV}$ in the numerical evaluations. Similar to that found in previous literature without considering the background field, the collisional energy loss at ${\cal Q} \neq 0$ grows quickly with increasing momentum, and a charm quark loses more energy as compared to the heavier bottom quark. This conclusion can be understood from the fact that the energy loss increases monotonically with increasing $v$ according to Fig.~\ref{elvsv}. Recall that for a given $p$, the incident quark with larger quark mass has a smaller velocity, and thus, one can naturally expect that the collisional energy loss of a charm quark is more pronounced as compared to a bottom quark carrying the same momentum. In Fig.~\ref{ravsp}, we also plot the energy loss ratio as a function of the heavy quark momentum. It can be shown that before it gets saturated at some large momentum $p$, the energy loss ratio has a quick increase in the small momentum region where the background field has a stronger impact on the suppression of $-dE/dx$ for a bottom quark. %Furthermore, for a charm quark, the quick increase of the ratio stops early as compared to a bottom quark. 
In principle, the observed $p$-dependence of the energy loss ratio can be also obtained from the corresponding $v$-dependence. Here, we should mention that an approximate constant ratio existing for $v \lesssim 0.5$ as shown in Fig.~\ref{elvsv} is not contradictory to the results given in Fig.~\ref{ravsp}. In fact, such an observation indicates that for a charm quark, the energy loss ratio is almost independent on the momentum up to $p \sim 0.8\,{\rm GeV}$, which becomes $p \sim 2.7\,{\rm GeV}$ for a bottom quark. Due to a rather narrow region, this behavior becomes invisible in Fig.~\ref{ravsp}. On the other hand, the saturation of the energy loss ratio in the large momentum region is actually related to the very weak $v$-dependence of the momentum when it gets large. Finally, it is worth pointing out that for a given temperature, although the energy loss ratio shows a dependence on the heavy quark velocity or momentum, its magnitude does not change dramatically when varying $p$ or $v$.

\begin{figure*}[t]
\begin{center}
\includegraphics[width=0.48\linewidth]{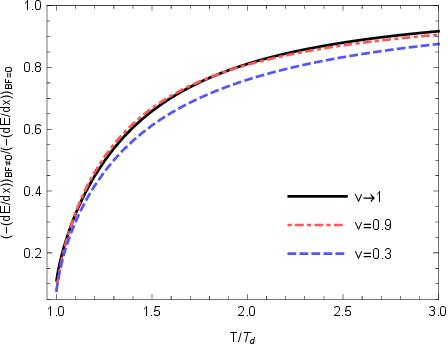}
\end{center}
\vspace{-5mm}
\caption{The ratio of the energy loss with and without the background field as a function of the temperature for different heavy quark velocities. The result in the ultrarelativistic limit $v\rightarrow 1$ comes from Ref.~\cite{Lin:2013efa} where the values of the background field are based on Eq.~(\ref{bf3}). }
\label{ravst}
\end{figure*}

As the background field strongly depends on the temperature, the most direct way to see the influence of the background field on the collisional energy loss is to study the energy loss ratio as a function of the temperature $T$. The corresponding results are shown in Fig.~\ref{ravst}, where we consider two different heavy quark velocities. Our results suggest that the background field suppresses $- d E/d x$ and the suppression is very significant in the entire semi-QGP region, from $T_d$ to about $2 \sim 3 T_d$. Especially, at the critical temperature, the heavy quark only losses $\sim 10\%$ of the energy that would be lost in the case where ${\cal Q}=0$. This can be attributed to the partial ionization of the plasma, which corresponds to a reduced number density of the thermal partons in the presence of a background field. On the other hand, the energy loss ratio increases quickly with the increasing temperature, and above the semi-QGP region, nonzero ${\cal Q}$ only shows a very weak influence on the heavy quark energy loss. This behavior can be understood by looking at the $T$ dependence of the background field. According to Eq.~(\ref{bf3}), the magnitude of the background field drops very quickly with increasing $T$ and becomes negligible at temperatures larger than $2 - 3 T_d$. In addition, there is only a moderate difference between the dashed curve for $v=0.3$ and the dot-dashed curve for $v=0.9$; therefore, the energy loss ratio has a weak dependence on the heavy quark velocity at any given temperature. This is actually consistent with our above findings.

We also compare our results with the energy loss ratio obtained in Ref.~\cite{Lin:2013efa}, where the authors computed the collisional energy in a semi-QGP with the perturbation theory. They focused on the ultrarelativistic limit, where $v \rightarrow 1$ and considered tree level Feynman diagrams with bare propagators. The infrared divergence associated with the $t$-channel contribution was regulated by choosing the Debye mass $m_D$ to be the cutoff, and thus, no resummation was considered for the soft momentum exchange. Consequently, the modification of the background field on the collisional energy loss turns out to be very simple, which is entirely encoded in the ${\sf q}$-dependent factor $(1 -3 {\sf q})^2$ for $SU(3)$. This is exactly the same as what we found based on the result of $-d E/d x$ in the weak coupling limit.\footnote{In the ultrarelativistic limit, the leading logarithmic contribution from $t$-channel is $\sim \ln (E T/m_D^2)$. In addition, the $u$-channel contribution cannot be neglected because it has a logarithmic enhancement $\sim \ln(E T/m_Q^2)$. However, in both cases, the modification due to the background field is simply given by $(1 -3 {\sf q})^2$.} The corresponding result in~\cite{Lin:2013efa} is also presented. However, in a perturbation theory, nonzero background field cannot be generated self-consistently from the equation of motion. Therefore, the values of the background field were extracted from the lattice simulations on the Polyakov loop in~\cite{Lin:2013efa}. For a direct comparison, the solid curve in Fig.~\ref{ravst} is obtained with the background field determined by Eq.~(\ref{bf3}). It is interesting to see an agreement between the results from \cite{Lin:2013efa} and our results at large velocities even for a realistic coupling constant. As a result, we can expect a smooth transition to the ultrarelativistic limit $v\rightarrow 1$, where in principle our result is no longer reliable due to the finite quark mass. In addition, ignoring the weak dependence on the heavy quark velocity, the energy loss ratio can be approximated by an overall factor $(1 -3 {\sf q})^2$ and according to our numerical results, the corresponding error is less than $10 \%$ when a realistic coupling constant is considered. 

\section{Summary and outlook}\label{summary}

In this work, by utilizing a background field effective theory, we calculated the collisional energy loss of a heavy quark propagating through a semi-QGP characterized by a nontrivial Polyakov loop. For temperatures close to the critical temperature, nonperturbative physics played an important role on the physical observables and such an effective theory, which well described the thermodynamics of the medium turned out to be an ideal tool to study the relevant problems in a semi-QGP. We considered the energy loss due to elastic scatterings between the incident heavy quark and the medium partons and showed that a nonzero background field, self-consistently generated from the equation of motion in the effective theory, had an important influence on both hard and soft contributions to the collisional energy loss. For hard scatterings, it was sufficient to use the bare propagator to compute the squared matrix element, while the thermal distribution functions of the medium partons were modified in the presence of a background field. For soft scatterings, it was necessary to take into account the HTL resummation in order to regulate the infrared divergence. Accordingly, the ${\cal Q}$-modified resummed gluon propagator which carried the correct screening information in the semi-QGP became the crucial ingredient to evaluate the energy loss associated with the soft processes.

Our results demonstrated that in the weak coupling limit, the collisional energy loss was independent on the momentum cutoff $q^*$, which has been introduced to separate the hard and soft contributions to $-dE/dx$. This was exactly the same as what happened at ${\cal Q}=0$, where the cutoff dependence was canceled between the hard and soft contributions. Furthermore, when ${\cal Q}\neq 0$, the cancellation did not rely on the explicit form of the background field, which was considered as a function of the temperature. It was also found that in the limit $g\rightarrow 0$, keeping only the leading logarithmic contribution $\sim \ln(1/g)$ in the collisional energy loss,  the background field modification on $-dE/d x$ became very simple and the total energy loss at ${\cal Q} \neq 0$ was just given by a ${\cal Q}$-dependent factor multiplied by the corresponding energy loss at ${\cal Q} = 0$. 

For a realistic coupling constant, on the other hand, the cutoff dependence existed in the collisional energy loss, and $-dE/dx$ had a notable dependence on $q^*$ when the coupling constant became large. We determined the cutoff by minimizing the total energy loss with respect to $q^*$, and thus, the corresponding $-dE/dx$ could be considered as a lower bound for the heavy quark energy loss. The numerical results suggested that although sharing some common properties with that at ${\cal Q}=0$, the collisional energy loss was reduced in the presence of a nonzero ${\cal Q}$. Although the suppression was not sensitive to the heavy quark velocity/momentum, it strongly depended on the temperature of the medium. Therefore, different from a simple $\sim T^2$ dependence at vanishing background field, $-d E/d x$ was expected to have a more involved $T$ dependence in a semi-QGP. In addition, the mass hierarchy of the energy loss, {\em i.e.}, a charm quark losing more energy than a bottom quark was also observed at nonzero ${\cal Q}$. With a given temperature, the background field led to a stronger suppression effect for a bottom quark although the magnitudes of the energy loss ratio for a charm and bottom quark did not differ significantly from each other. Finally, as the temperature approached to the critical temperature from above, the fast decrease of the energy loss ratio indicated a significant reduction of the collisional energy loss. Especially, the energy loss of a heavy quark was only $\sim 10 \%$ of that in a vanishing background field when $T=T_d$. However, the influence of the background field became negligible above $2-3 T_d$, where the saturated energy loss ratio got very close to one. Here, an interesting finding was that the simple form $(1-3{\sf q})^2$ describing the ${\cal Q}$ modification on $-dE/dx$ in the weak coupling limit could be approximately applied to a moderate coupling constant if we ignored the weak velocity dependence of the energy loss ratio.

The background field effective theory is a useful theoretical tool to study the physics in a semi-QGP, which is probably the most interesting region of the hot and dense medium being explored in high energy heavy-ion physics. Besides the heavy quark collisional energy loss discussed in this work, it can be also adopted to study other relevant observables, such as the photon and dilepton production, the radiated energy loss and so on. However, a full QCD analysis requires a nontrivial generalization of the effective theory where one needs to nonperturbatively include the contributions from the thermal fermions. Further work along this line needs to be carried out in the future.

\section*{Acknowledgements}
This work is supported by the NSFC of China under Project No. 12065004 and No. 12305135, by Guangxi Natural Science Foundation under Grant No. 2023GXNSFBA026027, by the promotion project of young college teachers in Guangxi province (Grant No. 2023KY0053), and by the Central Government Guidance Funds for Local Scientific and Technological Development, China (No. Guike ZY22096024).

%%%%%%%%%%%%%%%%%%%%%%%%%%%%%%%%%%%%%%%%%%%%%%%%%%%%%%%%%%
\bibliography{elossinBF}
\end{document}